# Visualizing Class Information

Haneen Abu Alfeilat
*Department Of Computer Science,
Mutah University, Al-Karak, Jordan*
haneenarafatabualfeilat@outlook.com

*Abstract:*
*A class is used in object oriented programming to describe each object in the system. It is as a template contains the methods and attributes for each object. The volume of information within the class has a role in the time required for its implementation, testing and understanding the class. Developers are dealing with large projects that contain a large number of Lines Of Code (LOC). So that, extracting information about the classes requires time and big effort from the developers. To solve this problem, we present a visualization approach to display class information for developers. Our method assumed each class is a cone chart and each cone represents one of the class metrics that include number of methods, attributes and the number of Lines Of Code (LOC). The height of each cone indicates the total number of methods, attributes and number of Lines Of Code (LOC) which found in a class. Also, each cone has a different color from the other to facilitate the distinction among them.*

*Keywords: class information, software visualization.*

## 1. Introduction

Class is a fundamental part of object oriented programming. It is a set of instructions to build a specific type of object. The attributes and methods for a particular object are defined inside the class. The size of a class is linked with the amount of information inside it. The class size can be measured by the number of attributes and by a number of methods.

The number of methods and their complexity is an important factor regard to developers. The larger number of methods that defined in class is required more effort to test and their implementation. Also, it makes class difficult to understand. The big classes need to divide to a smaller size. On the other hand, there are small classes need to merge with other classes in the system. Therefore, the developers need to know the number of methods for each class in a system.

Although, through maintenance phase developers can make changes on the consists of classes. For instance, the new method or attribute can be added for each class. Also, each method or attribute can be removed. Before attempting any change, the changes must be analyzed carefully by developers. So, the developers need to be aware about the size for each class in software.

The developers are facing millions of line codes with large software projects. Moreover, they deal with huge number of classes. As result, developers facing difficulty when reading source code to obtain on summary about the status of classes in a system. Reading and understanding the source code is a hard job, and requires time and effort from the developers.

In this paper, we propose a simple approach to visualize class information that includes the number of methods, attributes and number of Lines Of Code (LOC). Each class is represented as a cone chart. Each cone represents one of the components of the class, and each cone has a different color from the other. The height of each cone indicates a number of components in the class. The goal of our visualization is to add flexibility in viewing and comparing software classes in software projects. Thus, the developers obtain the information about the size of each class without requiring an in-depth analysis of the source code.

The visualizing class is not a new idea. The 2D visualization technique is used to display the internal structure of a class in [1], [2] and [3], named class blueprint. It displays the overall structure of a class, the relationship among the methods of the class, and how methods access attributes is. Although, some work has been done a 3D visualizing the class cohesion such as [4]. The angle 3D graph layout engine and XSLT transformations are used to select the component for visualizations. The cohesion is represented by the edges and the class is represented as a cylinder. Our visualization focuses in particular on the visualizing the size of the class.

The rest of the paper is organized as follows; Section 2 gives details of the proposed visualization. Areal example illustrates the proposed visualization presented at section3. Related work with our study in Section 4. Section 5 the limitations and future work, followed by conclusion in section 6.

## 2. The Proposed method

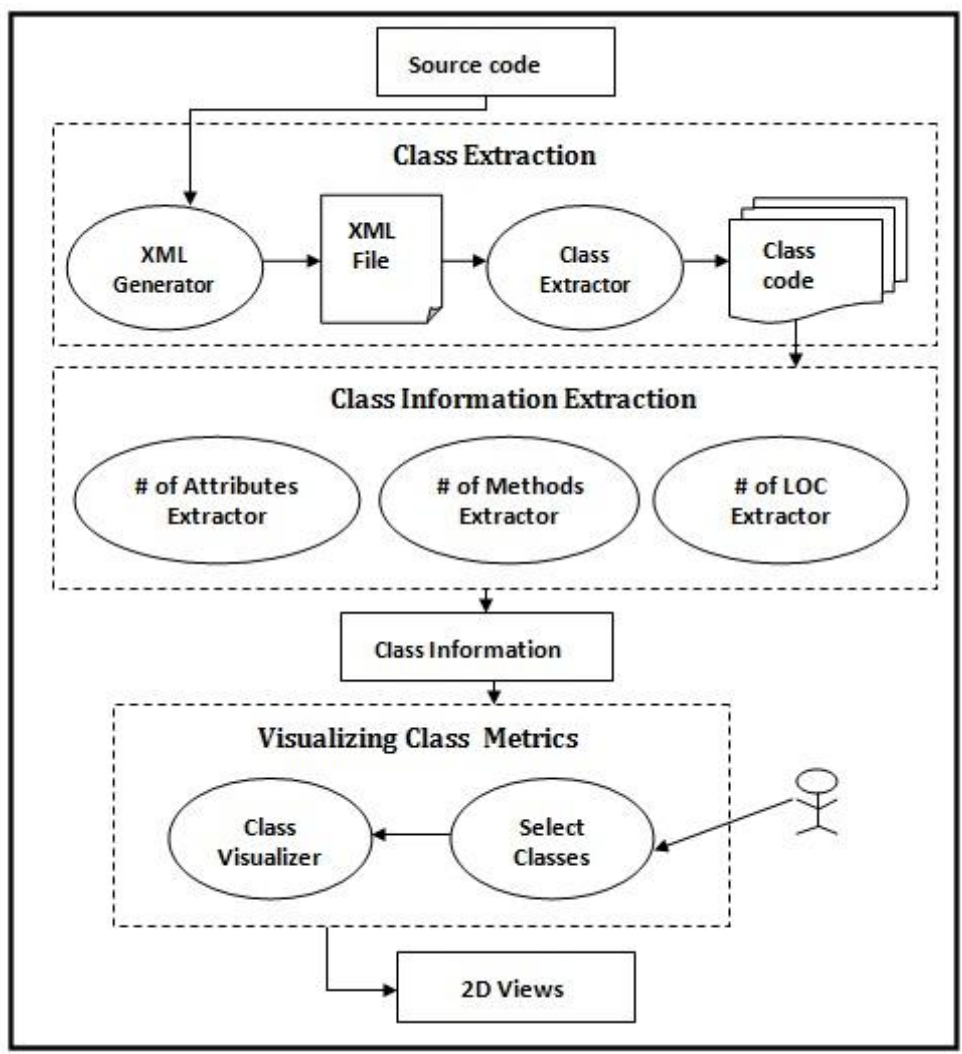

**Figure 1.** The main components for proposed visualization.

Figure 1 shows the main components of the proposed visualization. The proposed visualizations composed by three steps: Class Extraction, Class Information Extraction and Visulizing Class Metracies.Information extraction step focuses on extract classes from source code. Class Information Extraction includes three extractors to calculate the numbers for each component in the class. First extractor to calculate the number of methods. The second extractor to calculate number of attributes. Third extractor to calculate the number of Lines Of Code (LOC). Finally, the data is given to class visualizer to visualize the extracted data as 2D views.

### 2.1. Class Extraction

This component is responsible about extracting all classes' code from a source code of a project. This is done through two process, XML generator and class extractor. The XML generator is done by tool called scrMl. The tool converts the source code to XML file. After generated xml file, the classes are extracted by used XQuery to retrieve all classes in a project.

### 2.2. Class Information Extraction

This component is responsible about extracting all information for each class that include the number of methods, attributes and the number of Lines Of Code (LOC). This is done through three processes; each process is responsible about extract the total number for one of metrics.

Number of method extractor is responsible about extract the total number of methods in each class in a software project. It is done by two steps. First, the methods are retrieved through using the query to retrieve all methods for each class from XML file. Then, the number of methods is calculated automatically.

Number of Lines Of Code (LOC) extractor is responsible about extract the total number of Lines Of Code in each class in a software project. There are many tools to calculate the number of Lines Of Code(LOC). Of these tools, Code Counter Pro tool. It quickly counts several types of source code including XML type.

### 2.3. Visualizing Class Metrics

The aim of this process is to visualize the information for each class. It includes two process, class selector and class visualizer. The class selector done by user. The user selects the classes that want to get information about them. Then comes the role of class visualizer process. At this process, the class information that has been selected is given to visualize to display the data as 2D views. Each class visualized as a cone chart. The cone chart contains three cones. Each cone represents one of the components of the class; methods, attributes or lines of code. To distinguish between the cones, each cone has a different color .The green cone represents the method. The red cone represents the attribute. The Lines of code are represented as a blue cone. The height of each cone represents the total number of components in a class. Also, the total number is shown above each cone and class name is shown above a chart.

## 3. Real Example

We applied the proposed visualization on three classes defined in the package *jfreechart-1.0.5\source\org\org\jfree\chart*. The package is part of the open source java JFreeChart (http://www.jfree.org/jfreechart). The classes are *ChartColor*, *LegendRenderingOrder* and *PainMap* class.

Figure 2 shows the visualizations for the classes. Each class is visualized as a cone chart. Therefore, three charts are visualized in which class name is shown below each chart. The visualization provides useful information for the developers. For example, from Figure 3 we can observe the following information:

- *ChartColor* class has two methods and twenty five attributes. Also the number of lines of code in the class is seventy three.
- *LegendRenderingOrder* class has two methods, two attributes and the number of lines of code in the class is thirty one.
- *PaintMap* class has the largest number of methods in comparing to other classes. It has seven methods.

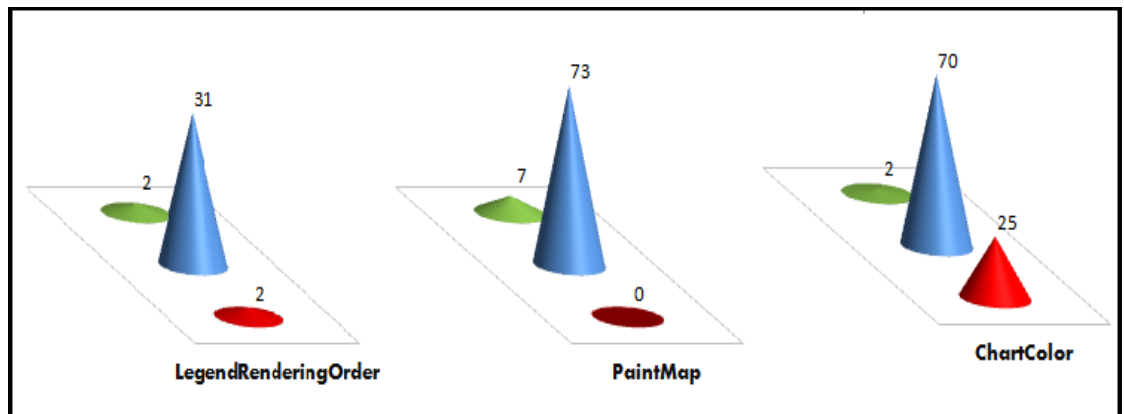

**Figure 2.** Visualization of class information for classes in Package in JFreeChart Project.

## 4. Related work

Many software visualization techniques have been published that focus on visualizing software evolution. For example, in [5], [6] present an approach to visualize the entire system evolution by using two-dimensional boxes and mapping the number of methods to width and the number of attributes to height. Also, each version of the software is represented as a column, and each class displayed as a line. The 2-D color maps are used to display several versions of software in [7]. Otherwise, Telea et al. [8] Presented the code flows visualization method that displays the source code line evolution across several file versions.

Visualizing packages, classes, and methods organization are issued many works. Harald et al.[9] proposed a 3D visual representations of object oriented programs. Each component in the system is represented as nodes and relationships between the components as edges. Object oriented software metrics also is represented in [10] as a solar system. The classes are displayed as blue planets and the orbit level represents the depth of a class. The other hand, the classes representing as boxes and packages as borders around the classes placed using a tree or sunburst layout in [11] using 3D visualizations.

Visualizing software system as a city is an idea introduced in several works, such as [12],[13]. Richard and Michele [12] use 3D visual to display object oriented software as a city metaphor. The classes are represented as buildings and the packages as districts, according to metrics such as the height and width. Also, Panas et al. [13] use a city metaphor to represent software. The city is named, "The Unified Single-

View City." The method is represented as a building, which is placed on blue plates that represented the classes in a system. The packages are displayed as green plates the height of a green plate depends on the depth of the package in the hierarchy. Mustafa [14] presented a visualization tool named BookViews. It represents a software project as a book . The book's chapters represent packages. The book's sections represent classes and its pages represent methods.

Several techniques for automatically summarizing source code have been addressed by researchers. For instance, Sonia et al. [15] present an approach to generate automatically descriptions as summarized for source code. The summary is a simple textual description to support comprehension for developers. Also, the source code is clustering in [16] through several steps. First, the latent semantic indexing is used to cluster source code to group that use similar words. Second, the groups are given the linguistic topics. Third, the topics are compared to identify links between them. Finally, visualization is used to illustrate how the groups are distributed over the system.

Mustafa and Adnan [17] proposed a framework to monitor the coupling of a class. The proposed framework extracts classes, and then calculates automatically a coupling for each class to visualize the results in useful views to developers.

Many tools are developed in software engineering. For example, Mustafa et al. [18] have built a lightweight monitoring tool called LIMOW. The tool aims to perform dynamic analysis for software. The tool is designed for wireless sensor software using the TinyOS platform. It provides flexibility and less complexity in the deploying environment for the developers. Another tool for extracting the software metrics from Java source code is developed in [19]. The tool is called EVOJAVA. The tool involves about the static analysis unlike tool in LIMOW. The aim of EVOJAVA is to preserve the identity of semantic software features (classes, methods, etc.) through sequential versions.

## 5. Limitations

In our approach, there are many class metrics that describe the structure of a class and relationship between classes did not visualize. Although, these metrics are useful in helping developers in the analysis of a classes in software project. From these metrics, for example, Number of Children (NOC), Depth of Inheritance Tree (DIT), Response for a Class (RFC), Number of Subunits (NUS) and other metrics. So that, in the future, we aim to visualize other classes matrices than the one used in this paper.

## 6. Conclusion

In this paper, we proposed an approach to visualize the number of methods, attributes and the number of Lines Of Code (LOC) that defined in each class in a software project. Visualization is useful for displaying the information of classes for developer without having to read the whole source code that is a difficult task for him. The approach assumes each class as chart cone, each cone represent one of the class metrics. The height of each cone linked with a total number for each component in the class. Also, each cone has a different color from the other.